\documentclass{emulateapj}
\usepackage{epstopdf}

\newcommand{\msun}{M$_{\sun}$}
\newcommand{\ldl}{$\lambda/{\Delta}{\lambda}$}
\newcommand{\logg}{$\log{g}$}
\newcommand{\teff}{T$_{eff}$}
\newcommand{\lbol}{$\log{{\rm L}_{bol}/{\rm L}_{\sun}}$}
\newcommand{\ki}{\ion{K}{1}}

\newcommand{\nai}{\ion{Na}{1}}
\newcommand{\rbi}{\ion{Rb}{1}}

\newcommand{\tii}{\ion{Ti}{1}}

\newcommand{\wat}{H$_2$O}

\newcommand{\hh}{H$_2$}
\newcommand{\kms}{km~s$^{-1}$}

\newcommand{\name}{SDSS~J125637.13$-$022452.4}
\newcommand{\namesh}{SDSS~J1256$-$0224}

\slugcomment{Submitted to ApJ 11 October 2008; Accepted 9 March 2009}

\shorttitle{L Subdwarf SDSS~J125637.13-022452.4}
\shortauthors{Burgasser et al.}

\begin{document}

\title{Optical and Near-Infrared Spectroscopy of the L Subdwarf SDSS~J125637.13-022452.4}

\author{Adam J.\ Burgasser\altaffilmark{1},
Soeren Witte\altaffilmark{2},
Christiane Helling\altaffilmark{3}, 
Robyn E.\ Sanderson\altaffilmark{1},
John J.\ Bochanski\altaffilmark{1},
and
Peter H.\ Hauschildt\altaffilmark{2}
}
\altaffiltext{1}{Massachusetts Institute of Technology, Kavli Institute for Astrophysics and Space Research, Building 37, Room 664B, 77 Massachusetts Avenue, Cambridge, MA 02139, USA; ajb@mit.edu}
\altaffiltext{2}{Hamburger Sternwarte, Gojenbergsweg 112, 21029 Hamburg, Germany}
\altaffiltext{3}{SUPA, School of Physics and Astronomy, University of St. Andrews, North Haugh, St. Andrews KY16 9SS, UK}

\begin{abstract}
Red optical and near-infrared spectroscopy are presented for
SDSS~J125637.13$-$022452.4, one of only four L subdwarfs reported to date.  
These data confirm the low-temperature,
metal-poor nature of this source, as indicated by prominent metal-hydride bands, alkali lines,
and collision-induced {\hh} absorption.  The optical and near-infrared spectra of SDSS~J1256$-$0224 are
similar to those of the sdL4 2MASS~J16262034+3925190, and we derive
a classification of sdL3.5 based on the preliminary scheme of Burgasser, Cruz, \& Kirkpatrick. 
The kinematics of SDSS~J1256$-$0224 
are consistent with membership in the Galactic inner halo, with estimated $UVW$ space velocities indicating a slightly prograde, eccentric and inclined Galactic orbit (3.5 $\lesssim$ $R$ $\lesssim$ 11~kpc; ${\mid}Z_{max}{\mid}$ = 7.5~kpc).
Comparison to synthetic spectra computed with the {\sc Phoenix} code,
including the recent implementation of kinetic condensate formation
({\sc Drift-Phoenix}),
indicate {\teff} $\approx$ 2100--2500~K and [M/H] $\approx$ $-$1.5 to $-$1.0 for {\logg} $\approx$ 5.0--5.5 (cgs), although there are clear discrepancies between model and observed spectra
particularly in the red optical region.  The stronger metal-oxide bands
present in the {\sc Drift-Phoenix} model spectra, a result of 
phase-non-equilibrium abundances of grain species, appears to contradict prior suggestions that grain formation is inhibited in metal-poor atmospheres; conclusive statements on the metallicity dependence of grain formation efficiency are as yet premature.   In addition, an apparent shift in the temperature scale of L subdwarfs relative to L dwarfs may obviate the need for modified grain chemistry to explain some of the former's unique spectral characteristics. 
\end{abstract}


\section{Introduction}

L subdwarfs are the lowest-luminosity and least-massive
halo population dwarf stars currently known \citep{mecs13}.
They derive their name from their gross spectral
similarities to the local L dwarf population of 
very low mass stars and brown dwarfs (see \citealt{kir05}), but
are distinguished by specific spectral
anomalies, including the presence of enhanced metal hydride absorption bands 
and unusually blue near-infrared colors.  These features are indicative of subsolar 
atmospheric abundances, as similar peculiarities distinguish 
metal-poor M subdwarfs from M dwarfs (e.g., \citealt{mou76,giz97,leg00}).
L subdwarfs, like M subdwarfs, exhibit kinematics consistent
with membership in the Galactic halo, with inclined and eccentric 
Galactic orbits indicating dynamical heating over long timescales and/or formation outside the Galactic disk
\citep{dah08,me0532plx,cus09}.
The low luminosities ({\lbol} $<$ $-3$)
and effective temperatures of L subdwarfs ({\teff} $<$ 3000~K; \citealt{leg00,rei06,me0532plx}),
coupled with their non-solar atmospheric abundances,
are of particular interest for studies of low temperature
atmospheres, testing thermochemistry and condensate formation processes in 
chemically peculiar environments (e.g., \citealt{ack01,lod02,hel06}).
In addition, as their inferred masses extend down to and below the
metallicity-dependent hydrogen burning minimum mass \citep{bur93,me0532},
L subdwarfs can potentially test metallicity dependencies on
low-mass star formation and brown dwarf evolution, relevant for instance in tracing
the terminus of the main sequence in globular clusters (e.g., \citealt{ric08}).

Despite their utility to atmospheric, star formation and Galactic population
studies, only four L subdwarfs have been reported to date:
the prototype 2MASS J05325346+8246465 \citep[hereafter 2MASS 0532+8246]{me0532},
2MASS~J16262034+3925190 \citep[hereafter 2MASS~J1626+3925]{me1626}, 2MASS~J06164006$-$6407194 \citep[hereafter 2MASS~J0616-6407]{cus09}, and 
{\name} \citep[hereafter {\namesh}]{siv09}.  The first three sources were identified serendipitously in the Two Micron All Sky Survey \citep[hereafter 2MASS]{skr06}, and
have been studied extensively at optical and near-infrared wavelengths
(e.g. \citealt{me0532plx,megmos,cus06,cus09,giz06,rei06,pat06,sch09,schil09}).
{\namesh} was found in the Sloan Digital Sky Survey \citep[hereafter SDSS]{yor00}
as part of a directed search for unusual red sources.
Both \citet{siv09} and \citet{sch09} have identified {\namesh} as an L subdwarf based on its optical spectrum (L-type, with unusually strong metal hydride bands),
blue near-infrared colors ($J-K_s = 0.10{\pm}0.03$; \citealt{schil09}),
and halo-like kinematics.   As yet, no detailed study of the optical and near-infrared spectral
properties of this unusual source have been made.

In this article, we present new observations of {\namesh} and conduct a detailed 
analysis of its spectral and kinematic properties.  Spectroscopic observations spanning the red optical and near-infrared are described in $\S$~2.
The empirical properties of 
{\namesh} are assessed in $\S$~3, including classification (on the preliminary
scheme of \citealt{megmos}), 
distance estimation, kinematics and Galactic orbit.  In $\S$~4 we examine the 
atmospheric properties
of this source by comparing its colors and spectra 
to the latest generation of the {\sc
Cond-Phoenix} atmosphere simulations \citep{hau97} and to the {\sc
Drift-Phoenix} model atmospheres \citep{deh07,hel08b,wit08}, the latter of
which includes a kinetic approach of phase-non-equilibrium
dust formation. Results are summarized in $\S$~5.
 
\section{Observations}

\subsection{Red Optical Spectroscopy}

Optical spectra of {\namesh} were obtained
on 2006 May 7 (UT) using the Low Dispersion Survey Spectrograph (LDSS-3)
mounted on the Magellan 6.5m Clay Telescope.
LDSS-3 is an imaging spectrograph, upgraded from the original
LDSS-2 \citep{all94} for improved red sensitivity.
Conditions during the observations were
clear with excellent seeing (0$\farcs$6 at $i^{\prime}$-band).
The VPH-red grism (660 lines/mm) with a 0$\farcs$75 wide
(4 pixels) longslit mask was used, with the slit
aligned to the parallactic angle.  This configuration provides
6050--10500 {\AA} spectra across the entire chip
with an average resolution of {\ldl} $\approx$ 1800 and dispersion
along the chip of $\sim$1.2~{\AA}/pixel.  The OG590 longpass filter
was used to eliminate second order light shortward of 6000~{\AA}.
Two slow-read exposures of 750~s each were obtained at an airmass
of 1.12. We also observed the
G2~V star G 104-335 ($V$ = 11.7)
immediately after the {\namesh} observation and
at a similar airmass (1.14) for telluric absorption correction.  
The flux standard LTT 7987 
(a.k.a.\ GJ 2147; \citealt{ham94}) was observed 
on the same night using an identical slit 
and grism combination. All spectral observations 
were accompanied by HeNeAr arc lamp and flat-field
quartz lamp exposures for dispersion and pixel response calibration.

LDSS-3 data were reduced in the IRAF\footnote{IRAF is 
distributed by the National Optical
Astronomy Observatories, which are operated by the Association of
Universities for Research in Astronomy, Inc., under cooperative
agreement with the National Science Foundation.} environment \citep{tod86}. 
Raw images were first corrected for amplifier bias voltage, 
stitched together, and subtracted by a median-combined
set of slow-read bias frames taken during the afternoon.  
These processed images were then divided by a
median-combined, bias-subtracted and normalized
set of flat field frames.
The LTT 7987 and G 104-335 spectra were optimally
extracted first using the APALL task with background subtraction.  
The spectrum of {\namesh} was then extracted using
the G star dispersion trace as a template.  Dispersion solutions were
determined from arc lamp spectra extracted 
using the same dispersion trace; 
solutions were accurate to $\sim$0.08 pixels, or $\sim$0.1~{\AA}.  
Flux calibration (instrumental response correction) was determined
using the tasks STANDARD and SENSFUNC with
observations of LTT 7987, which we have found provide sufficient
calibration to $<$10\% over the 6000--9000 {\AA} spectral band \citep{megmos}.
Corrections to telluric O$_2$ (6855--6955 {\AA} B-band,
7580--7740 {\AA} A-band)
and H$_2$O (7160--7340 {\AA}, 8125--8350~{\AA}, 9270--9680 {\AA}) absorption bands
were determined by linearly interpolating over these features in the 
G dwarf spectrum, dividing by the uncorrected spectrum, 
and multiplying the result with the spectrum of {\namesh}.  
The two spectra of {\namesh} were then coadded to improve 
signal-to-noise, which ranged from $\sim$15 at the 6600~{\AA} peak
to a maximum of $\sim$45 at 8500{\AA}.

\begin{figure}
\centering
\epsscale{1.0}
\plotone{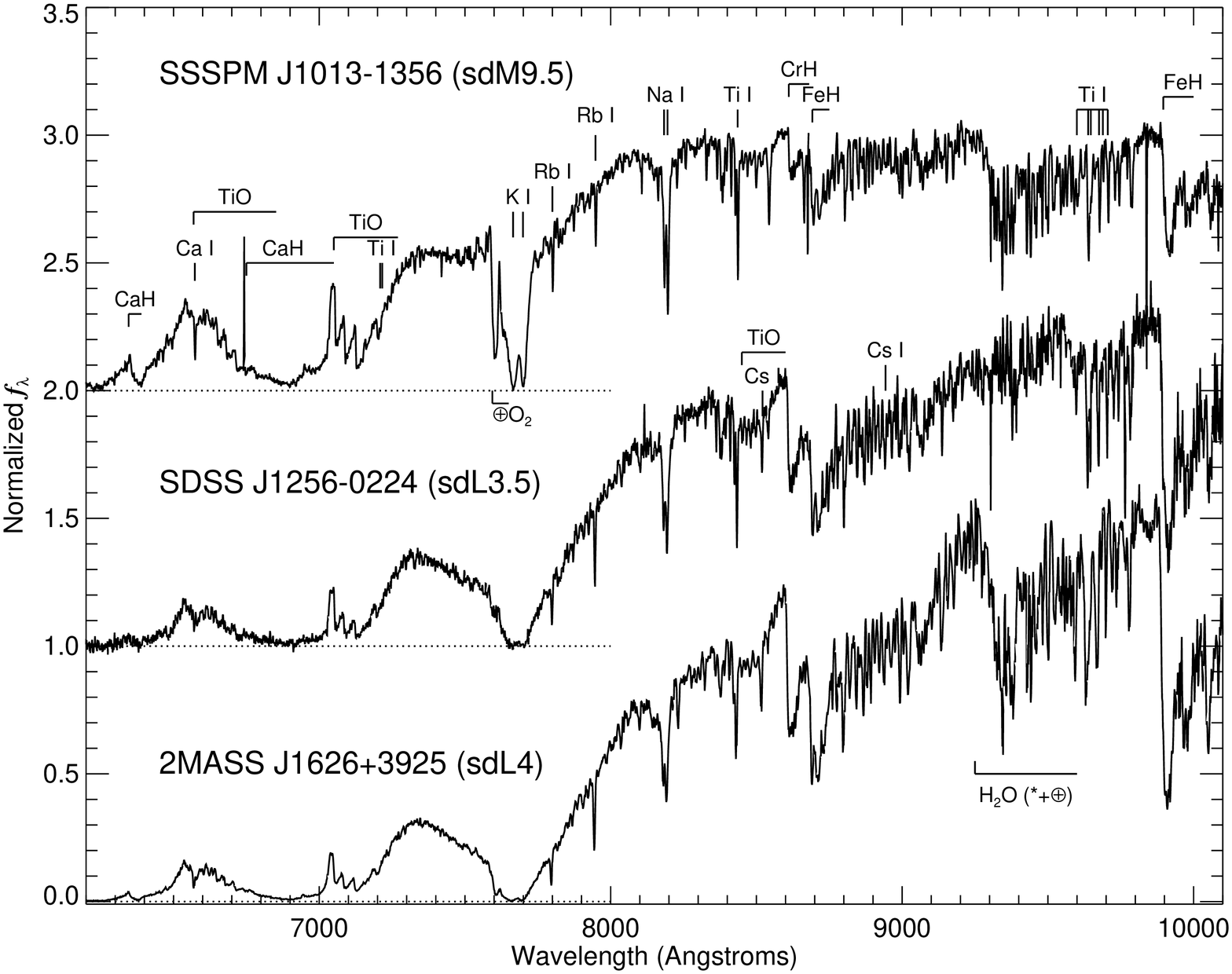}
\caption{Red optical spectra of the SSSPM~J1013-1356 (top), {\namesh} (middle)
and 2MASS~J1626+3925 (bottom).  Spectra are normalized in the 8500--8600~{\AA}
region and offset for comparison (dotted lines).  Primary spectral
features are indicated, as well as regions of strong telluric absorption
($\oplus$) in the spectra of SSSPM~J1013-1356 and 2MASS~J1626+3925. 
\label{fig_optspec}}
\end{figure}

The reduced red optical spectrum of {\namesh}
is shown in Figure~\ref{fig_optspec}, compared to equivalent data
for the sdM9.5 SSSPM~J1013-1356 \citep{sch1013} and
2MASS~J1626+3925 \citep{megmos}.\footnote{These data were
obtained with the Gemini Multi-Object Spectrometer \citep{hoo04}.}
Our data for {\namesh} have considerably higher signal-to-noise 
than the original SDSS discovery spectrum \citep{siv09} and higher resolution
than contemporaneous observations by \citet{sch09}.
As originally pointed out by \citet{siv09}, the optical spectrum of {\namesh} 
exhibits several characteristics indicative of an L dwarf, 
including an overall red spectral slope from 6000--8500~{\AA};
strong molecular bands of CrH (8600~{\AA}) and FeH (8700 and 9900~{\AA}); and
alkali line absorption from \ion{K}{1} (7700~{\AA} doublet),
\ion{Na}{1} (8182/8193~{\AA} doublet), \ion{Rb}{1} (7798 and 7946~{\AA}) and \ion{Cs}{1} (8519 and  possibly 8941~{\AA}).  Equivalent width (EW) measurements for these lines
are listed in Table~\ref{tab_ews}.
The \ion{K}{1} doublet is extremely broadened, producing a V-shape notch in the spectrum
that spans 7300--8100~{\AA}, also characteristic of L dwarf spectra.
There are a number of peculiar features in the spectrum of {\namesh} that
are not common to L dwarf spectra, however, 
including unusually strong bands of CaH (6900~{\AA}) and TiO (7200 and 8400~{\AA}),
and numerous metal lines from \ion{Ca}{1} (6571~{\AA}), 
\ion{Ca}{2} (8541~{\AA}) 
and \ion{Ti}{1} (7204, 8433 and 9600--9700~{\AA}\footnote{Absorption in the 9600--9700~{\AA}
spectrum of 2MASS~J1626+3925 was incorrectly associated with TiH
by \citet{me1626}.  
\citet{cus06} and \citet{rei06} have since identified these features as 
arising from the a $^5$F--z $^5$F$^o$ multiplet  of {\tii}, and we adopt
these identifications here.}).
The absence of these species in L dwarf spectra is largely attributed
to the formation of Ca-Ti and Ca-Al mineral condensates 
(e.g., \citealt{all01,lod02,hel08a}),
and their presence in L subdwarf spectra has been interpreted as
an indication of inhibited condensate formation \citep{me0532,rei06}.
Whether or not this is an accurate interpretation (see $\S$~4.3), 
unsually strong CaH and metal-line absorption is a characteristic trait of 
metal-poor M subdwarf spectra (e.g., \citealt{mou76,giz97}) and the presence of these
features in the spectrum of {\namesh} supports its characterization as a metal-poor, 
low-temperature dwarf.

Specific comparison of {\namesh}
to 2MASS~J1626+3925
reveals remarkable similarities between these two sources, although
the latter exhibits
somewhat stronger FeH, CrH,  \ion{Rb}{1} and \ion{Cs}{1} absorption features
and somewhat weaker  \ion{Na}{1} lines.  Indeed, the variation in
features between the three spectra shown in 
Figure~\ref{fig_optspec} suggests a sequence of very late-type,
metal-poor subdwarfs, with {\namesh} having intermediate 
line and band strengths.
The specific spectral classification of {\namesh} is discussed in 
detail in $\S$~3.1.
Note that no significant H$\alpha$ emission or absorption is detected
in any of these spectra.

\subsection{Near-Infrared Spectroscopy}

Low resolution near-infrared
spectral data for {\namesh} were obtained in clear conditions
on 2005 March 23 (UT) using the SpeX spectrograph \citep{ray03}
mounted on the 3m NASA Infrared Telescope Facility (IRTF).
We used the prism-dispersed mode of SpeX with a 0$\farcs$5
slit (aligned to the parallactic angle), providing 0.75--2.5~$\micron$
spectroscopy with resolution {\ldl} $\approx 120$
and dispersion across the chip of 20--30~{\AA}~pixel$^{-1}$.
{\namesh} was observed at an airmass of 1.08.
Four exposures of 180~s each were obtained 
in an ABBA dither pattern along the slit.
The A0~V star HD~111744 was observed immediately
before {\namesh} at a similar airmass (1.07)
for telluric absorption and flux calibration.
Internal flat field and Ar arc lamps were observed
with the target and calibrator source
for pixel response and wavelength calibration.

Data were reduced using the SpeXtool package, version 3.1
\citep{cus04} using standard settings. Raw science images were first
corrected for linearity, pair-wise subtracted, and divided by the
corresponding median-combined flat field image.  Spectra were optimally extracted using the
default settings for aperture and background source regions, and wavelength calibration
was determined from arc lamp and sky emission lines.  The multiple
spectral observations were then median-combined after scaling individual
spectra to match the highest signal-to-noise
observation.  Telluric and instrumental response corrections for the science data were determined
using the method outlined
in \citet{vac03}, with line shape kernels derived from the arc lines. 
Adjustments were made to the telluric spectra to compensate
for differing \ion{H}{1} line strengths and pseudo-velocity shifts.
Final calibration was made by
multiplying the spectrum of {\namesh} by the telluric correction spectrum,
typically accurate to within 10\% across the 0.8--2.5~$\micron$ window
(see \citealt{metgrav}).
Instrumental response was determined through the ratio of the observed A0~V spectrum
to a scaled, shifted and deconvolved Kurucz\footnote{See \url{http://kurucz.harvard.edu/stars.html}.} model spectrum of Vega. Signal-to-noise ranged from $\sim$80 at the $J$-band peak ($\sim$1~$\micron$) to $\sim$15 at $K$-band
($\sim$2.2~$\micron$).

\begin{figure}
\centering
\epsscale{1.0}
\plotone{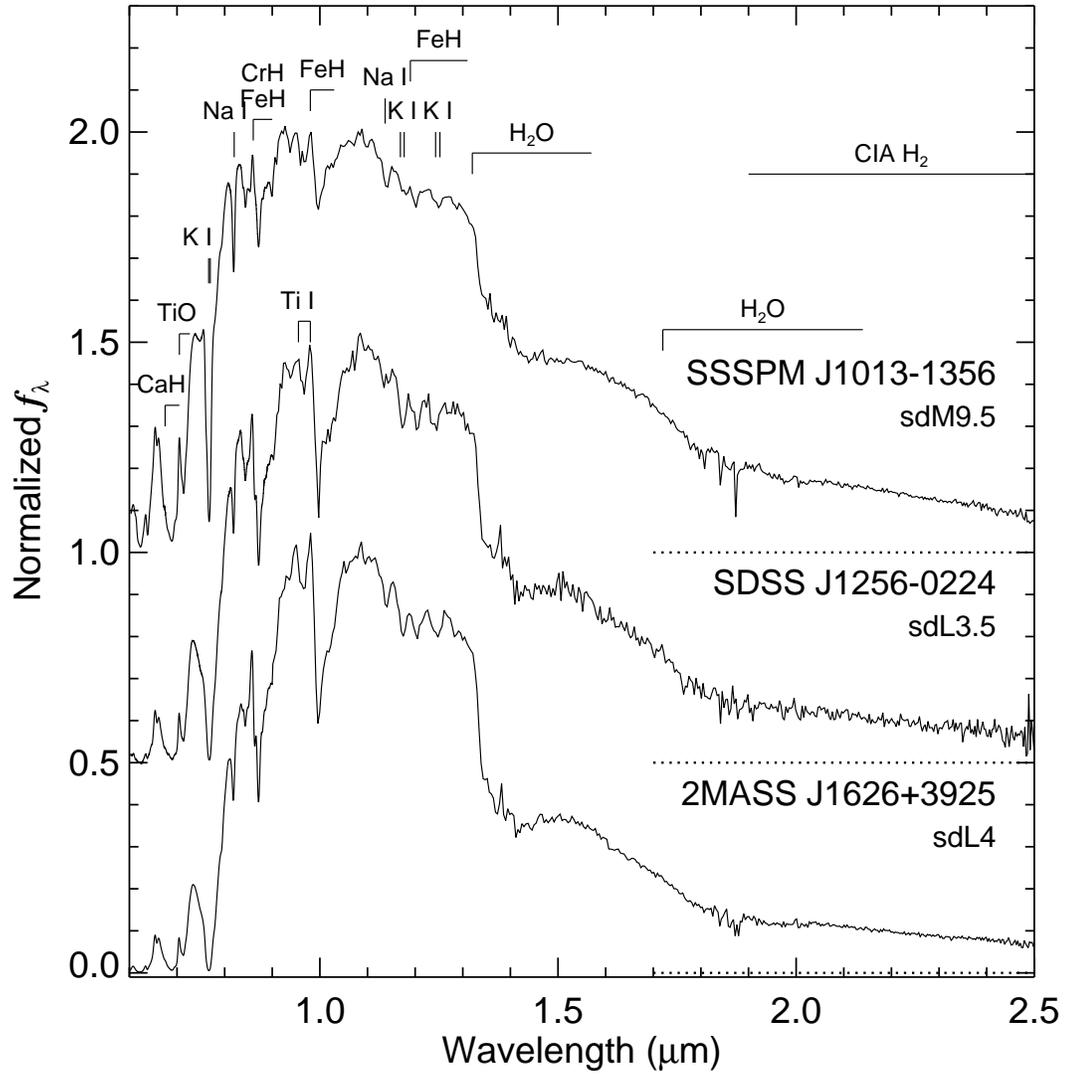}
\caption{Low resolution near-infrared spectra of SSSPM~J1013-1356 (top), {\namesh} (middle)
and 2MASS~J1626+3925 (bottom), all obtained with SpeX in prism mode.
Joined to these are the optical data from Figure~\ref{fig_optspec},
smoothed to the same resolution ({\ldl} $\sim$ 120).
Spectra are normalized in the 0.9--1.0~$\micron$ region and offset
for comparison (dotted lines).  Primary spectral features are indicated.
\label{fig_nirspec}}
\end{figure}

The reduced near-infrared spectrum of {\namesh} is shown in Figure~\ref{fig_nirspec},
again compared to equivalent SpeX prism data for
SSSPM~J1013-1356 and 2MASS~J1626+3925
\citep{me1626}.
All three spectra show similar overall
spectral morphologies, with strong molecular absorption features
and relatively blue 1.0--2.5~$\micron$ spectral slopes.
Again, 2MASS~J1626+3925 appears to be 
most similar to {\namesh} in terms of detailed spectral features.
The strong FeH band at 0.99~$\micron$ present in the 
optical spectrum of {\namesh} is clearly discerned in these data, intermediate in
strength between  SSSPM~J1013-1356 and 2MASS~J1626+3925 but considerably
stronger than any normal L dwarf.
Strong FeH absorption
is also present in the 1.2--1.3~$\micron$ region.  {\wat} absorption
is quite pronounced at 1.4~$\micron$, but the 1.8~$\micron$ band is notably weaker.
CO absorption at 2.3~$\micron$, a hallmark of M and L dwarf near-infrared spectra,
is absent in the three subdwarf spectra shown here.
The weakened 1.8~$\micron$ {\wat} and  2.3~$\micron$ CO bands, and the overall
blue spectral slope, 
can be attributed to enhanced H$_2$ absorption in all three sources, peaking
at 2.1~$\micron$ but spanning much of the 1.5--2.5~$\micron$ region shown \citep{lin69,sau94,bor97}.
This absorption produces the very blue near-infrared color of {\namesh},
$J-K_s$ = 0.10$\pm$0.03 \citep{schil09}\footnote{Note that the 2MASS $J-K_s$ = 0.66 color reported by \citet{siv09} is in fact an upper limit; this source was not detected by 2MASS in the $K_s$ band.}, quite
distinct from the colors of normal L dwarfs ($J-K_s$ $\approx$ 1.5--2.5, \citealt{kir00}).
Atomic lines arising from the neutral alkalis \ion{Na}{1}
and \ion{K}{1} are also seen, but the resolution of the SpeX prism data
is insufficient to obtain meaningful EW measurements.

\section{Characterization of {\namesh}}

\subsection{Spectral Classification}

\begin{figure}
\centering
\epsscale{1.0}
\plotone{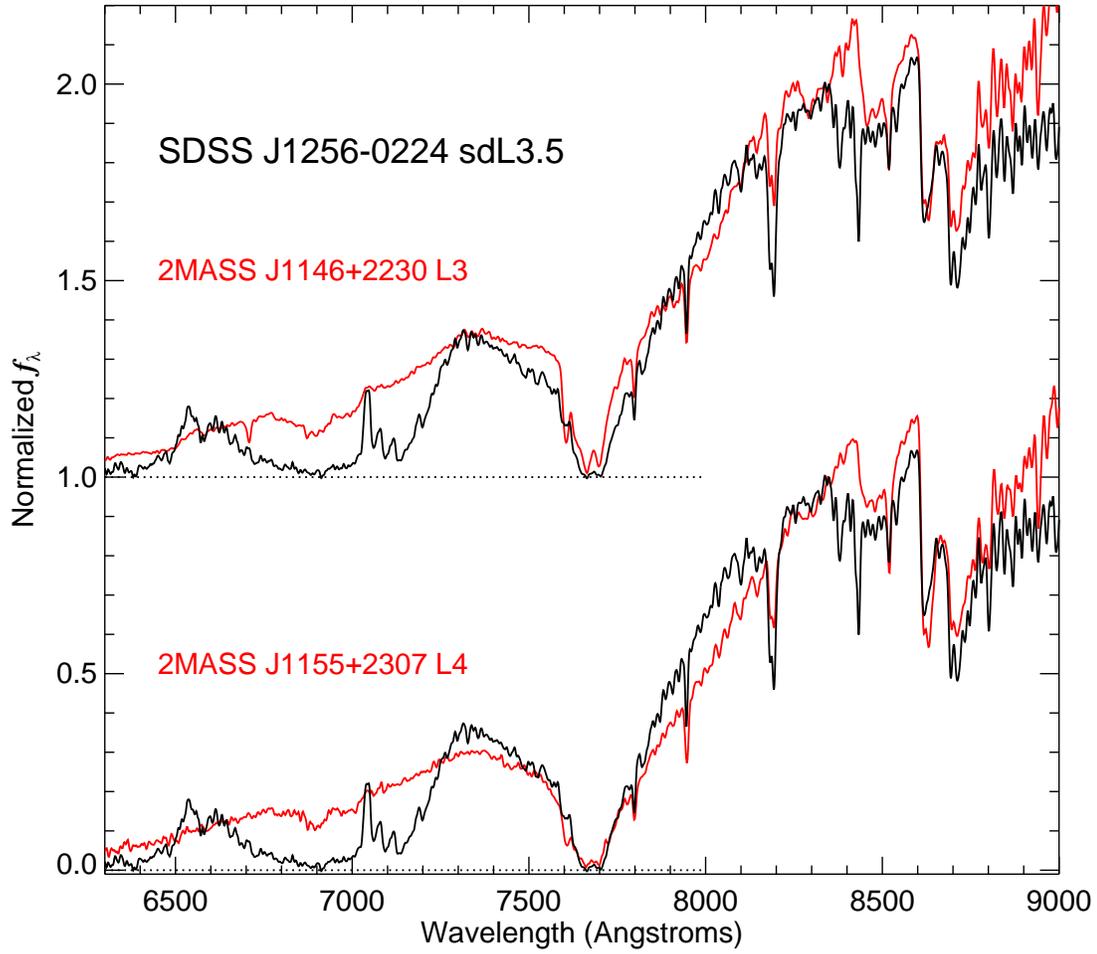}
\caption{Optical spectrum of {\namesh} (black lines) compared to the L dwarf optical
standards (red lines) 2MASS~J1146+2230 (L3) and 2MASS~J1155+2307 (L4).  All spectra
are normalized in the 8250--8350~{\AA} and offset for clarity (dotted lines).
In the 7300--9000~{\AA} range, the spectral morphology of {\namesh} is intermediate between the two standards, indicating an sdL3.5 spectral type for this source based on the scheme
of \citet{megmos}.
\label{fig_optclass}}
\end{figure}

The most basic characterization of a late-type dwarf like {\namesh} is its spectral
classification.  However, while a well-defined red optical classification scheme exist for L dwarfs (\citealt{kir99,kir00}; see also \citealt{geb02} 
for discussion on the near-infrared classification of L dwarfs), there are simply too few L
subdwarfs to define a robust scheme.  We therefore followed the approach outlined
in \citet{megmos}, comparing the 7300--9000~{\AA}
spectrum of {\namesh} to equivalent-resolution spectra of the
L dwarf spectral standards defined in \citet{kir99}.  
Figure~\ref{fig_optclass} shows the two 
best-matching standards, the L3 2MASS~J11463449+2230527
and the L4 2MASS~J11550087+2307058 (data from \citealt{kir99}\footnote{These data
were obtained with the Low Resolution Imaging Spectrograph \citep{oke95}.}). Focusing on the
7300--9000~{\AA} region, the spectrum of {\namesh} appears to lie intermediate between these two
standards, based on the height of the 7300~{\AA} spectral peak; the
depth and breadth of the 7700~{\AA} {\ki} doublet; and the
depts of the 8500~{\AA} TiO, 8600~{\AA} CrH and 8700~{\AA} FeH
bands.  There are discrepancies between {\namesh}
and the L dwarf standards in this region, notably
the spectral slope between 7800~{\AA} and 8400~{\AA}, the depth of
the 8200~{\AA} {\nai} doublet (stronger in {\namesh}), and the presence of additional
\ion{Ca}{1}, \ion{Ca}{2} and \ion{Ti}{1} lines in the spectrum of {\namesh}.
However, these differences are not nearly as extreme as those
at shorter (i.e., the CaH and TiO bands
between 6600--7300~{\AA}) or longer wavelengths (i.e., the strong H$_2$ absorption).
The comparisons shown in Figure~\ref{fig_optclass} indicate
a spectral type of sdL3.5 for {\namesh} on the \citet{megmos} scheme.  This is only 0.5 subtypes
earlier than the sdL4 classification of 2MASS~J1626+3925, 
consistent with both
the overall similarities between the spectra of these two sources 
and slight differences in their atomic line and molecular band strengths.

\subsection{Absolute Brightness and Distance}

Parallax distance measurements have recently become available for low-temperature subdwarfs spanning types sdM7 to sdL7
\citep{mon92,dah08,me0532plx,schil09}, including {\namesh} for which
\citet{schil09} determine $d$ = 90$\pm$23~pc.
This measurement is fairly uncertain, so we have compared it to the
linear absolute magnitude/spectral type relations recently
quantified by \citet{cus09} for ultracool subdwarfs:
\begin{equation}
M_J = 8.02+0.313\times{SpT} 
\end{equation}
\begin{equation}
M_H = 7.77+0.300\times{SpT} 
\end{equation}
\begin{equation}
M_{K_s} = 7.44+0.320\times{SpT} 
\end{equation}
where SpT(sdM7) = 7, SpT(sdL0) = 10, etc.  These relation predict
$M_J$ = 12.23$\pm$0.22, $M_H$ = 11.81$\pm$0.23 and $M_{K_s}$ = 11.75$\pm$0.25 for a type of sdL3.5$\pm$0.5,
and hence $d$ = 66$\pm$9~pc for {\namesh} based on the photometry of \citet{schil09} (the distance uncertainty includes contributions from photometry, a 0.5 subtype classification
uncertainty, covariance matrix elements for the \citealt{cus09} relations, and variation in distance estimates between $JHK_s$ values).  This distance inferred estimate is considerably closer than the mean value from \citet{schil09} but nevertheless consistent within experimental uncertainties.  It is also roughly half the 120~pc estimate of \citet{megmos} and 50\% larger than the 42~pc estimate of \citet{siv09}.  As the \citet{cus09} relations provide the
most accurate assessment of the ultracool subdwarf spectral type/absolute magnitude scale thus far, we use our estimated value from these relations for subsequent analysis.

\subsection{Kinematics}
 
The proper motion of {\namesh} as measured by \citet{schil09}, in conjunction with
our estimated distance, yields a tangential velocity
$V_{tan}$ = 186$\pm$26~{\kms}, where
the uncertainty is dominated by the distance estimate.  Note that this value is again
roughly half that estimated by \citet{megmos} due to the reduction in the estimated distance.
Nevertheless, this motion is
still indicative of halo kinematics.
A radial velocity ($V_r$) for {\namesh} was computed from its optical spectrum, 
by comparing the measured line centers of the atomic lines listed in Table~\ref{tab_ews}
to vacuum wavelengths obtained from the National Institute of Standards and Technologies
atomic line database\footnote{\url{http://physics.nist.gov/asd3}} \citep{ral08}.
A heliocentric Doppler shift of $-$130$\pm$11~{\kms} was determined, which includes 
a barycentric motion correction of $-$15~{\kms}.  The uncertainty in $V_r$
includes the standard deviation in the line centers and 
a systematic uncertainty of 4~{\kms} based on the uncertainty of the wavelength dispersion solution.  
This value is nominally consistent with the $-$90$\pm$40~{\kms}
cross-correlation measurement of \citet{siv09}; our 
improvement in precision is likely due to our higher signal-to-noise spectral data.

Adding in our radial velocity measurements, we find space velocities
[$U,V,W$] = [$-115{\pm}11$, $-101{\pm}18$, $-150{\pm}9$]~{\kms} in
the Local Standard of Rest (LSR), assuming an LSR solar motion of 
[$U,V,W$]$_{\sun}$ = [10, 5.25, 7.17]~{\kms} \citep{deh98}.
Note that we adopt a right-handed velocity coordinate system with positive 
$U$ pointing radially inward toward the Galactic center.
The large space motions in all three LSR components are again strong indications of Galactic halo membership for {\namesh}.

\subsection{Galactic Orbit}
 
To explore the kinematics of {\namesh} is more detail, we calculated
its Galactic orbit using the inferred $UVW$ velocities as initial conditions.
These velocities were first converted to a 
Galactic inertial frame (assuming $V_{LSR}$ = +220~{\kms}; \citealt{ker86}),
and the position and distance of {\namesh} transformed to
galactocentric rectangular coordinates [$X,Y,Z$] aligned with [$U,V,W$],
assuming a Solar position
of [$-$8.5, 0, 0.027]~kpc \citep{ker86,che01}.  We adopt
the convention of positive $X$ pointing toward the Galactic center
to align with our definition for $U$.  
The Galaxy was modeled using a set of static potentials comprising a spherically-symmetric halo and bulge and an axisymmetric, thin exponential disk.  This model is a simplified version of the one described in \citet{deh98} and includes the three density distributions:
\begin{eqnarray*}
\rho_{\mathrm{bulge}}(r) &=& \rho_{b0} \left(\frac{r}{a_b}\right)^{-\alpha_b}\\
\rho_{\mathrm{halo}}(r) &=& \rho_{h0}\left(\frac{r}{a_h}\right)^{-\alpha_h}\left(1-\frac{r}{a_h}\right)^{\alpha_h-\beta_h}\\ 
\rho_{\mathrm{disk}}(R,z) &=& \Sigma_0 e^{-R/R_d} \delta(z)
\end{eqnarray*}
with spherical and polar coordinates $r \equiv \sqrt{X^2 + Y^2 + Z^2}$ and $R \equiv \sqrt{X^2 + Y^2}$.  With these simplifications, the potentials corresponding to the bulge and halo densities can be expressed in terms of special functions, and the potential of the disk evaluated on a grid by  numerical integration.  The parameters $\rho_{b0}, a_b, \alpha_b, \rho_{h0}, a_h, \alpha_h, \beta_h, \Sigma_0, R_d$ were adopted as given for Models I and II in Table 2.3 of \citet{bin08}, which fit the measured rotation curve of the Galaxy but bracket the range of allowable disk/halo mass ratio in the Solar circle (Model I describes a Galaxy dominated by the disk mass at the solar circle, in Model II the halo mass dominates at the solar circle).  
The orbit of {\namesh} was integrated using a second-order leapfrog method (kick-drift-kick) with a constant timestep of 1 kyr for a total simulation time of 1 Gyr centered on the present epoch.  Energy was conserved to better than 1 part in $10^{-4}$ over the full length of the simulation, with the error dominated by the resolution of the grid on which the disk force and potential were interpolated.  The $Z$ component of angular momentum was conserved to 1 part in $10^{-13}$.

\begin{figure}
\centering
\epsscale{1.0}
\plotone{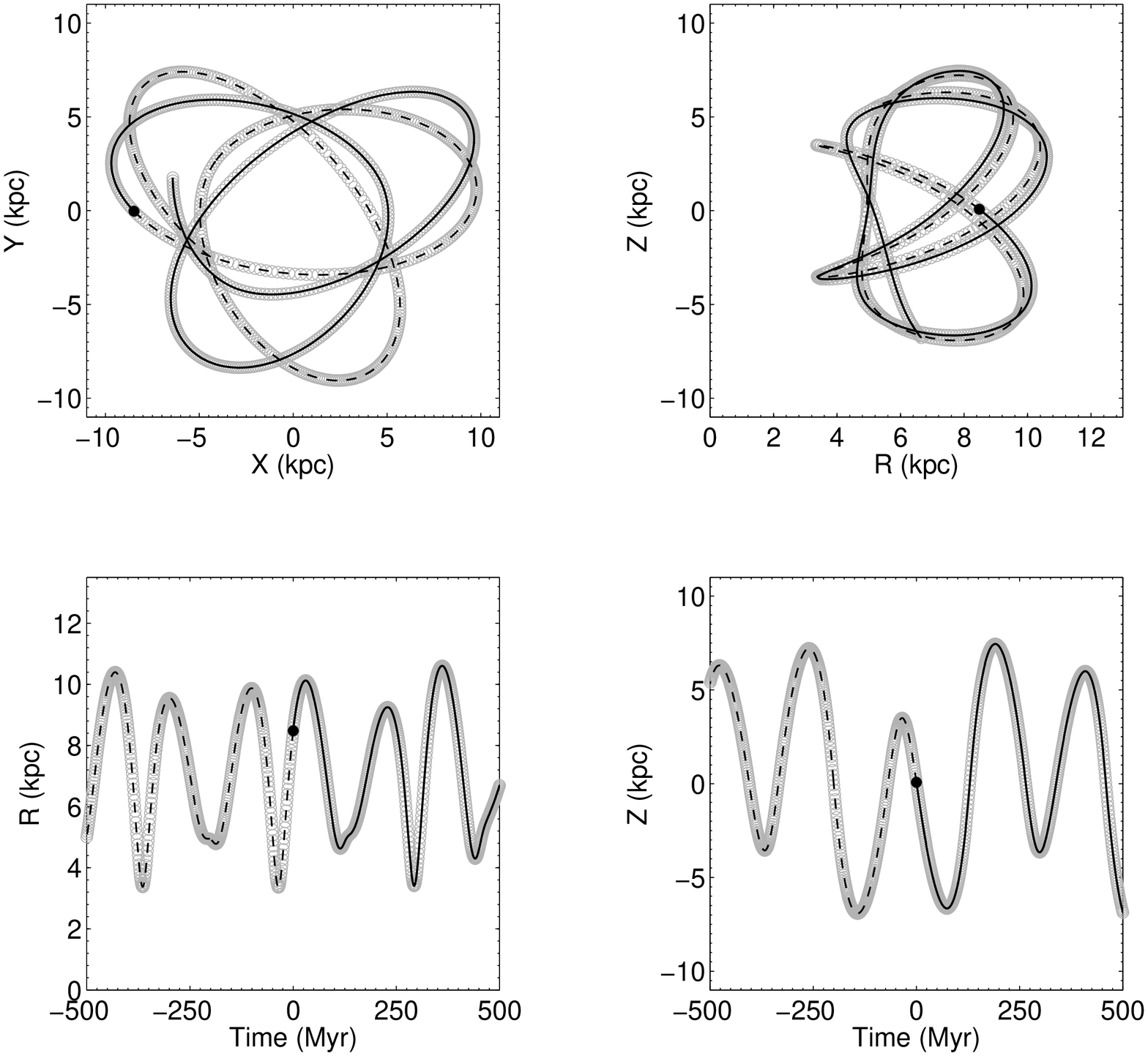}
\caption{Galactic orbit of {\namesh} over 1~Gyr centered on the current epoch, based on Galactic Model I from \citet{bin08}.
Upper left and right panels shows orbit in [$X,Y$] and [$R,Z$] inertial frame coordinates; the current position of the Sun ($X_{\sun}$ = -8.5~kpc,
$Z_{\sun}$ = +27~pc; \citealt{ker86,che01}) is indicated by the ${\sun}$ symbol.
Bottom panels show time evolution of $R$ and $Z$.
In all panels, past motion is indicated by dashed lines, future motion by solid lines, and current position by the black point.  
\label{fig_orbit}}
\end{figure}

Figure~\ref{fig_orbit} displays the resulting orbit of {\namesh} using Model I (Model II provides essentially identical results).  
The general character of this orbit is similar to several other ultracool subdwarfs, with a prograde eccentric orbit with apoaps near the Solar radius (3.5 $\lesssim$ R $\lesssim$ 11~kpc, $e$ = 0.5)
and substantial deviations from the Galactic plane 
($Z_{max} \approx {\pm}$7.5~kpc).  In terms of eccentricity, the orbit of {\namesh} is more similar to that of 2MASS~J0532+8246 ($e$ = 0.5, 3 $\lesssim$ R $\lesssim$ 8.5; \citealt{me0532plx}) and less ballistic than that of the sdM8 LSR~1425+7102 \citep{lep1425} which plunges to within 1~kpc from the Galactic center \citep{dah08}.  The inclination of {\namesh}'s orbit, $\tan{i} \approx R_{max}/Z_{max}$, is quite a bit larger, nearly 45$\degr$ with respect to the Galactic plane as compared to $\sim$15$\degr$ and $\sim$30$\degr$ for 2MASS~J0532+8246 and LSR~1425+7102, respectively.  

The orbital characteristics of {\namesh}, 2MASS~J0532+8246 and LSR~1425+7102 are 
all consistent with membership in the Galaxy's inner halo
population \citep{chi01,car07}.
The inner halo is believed to dominate the halo population in the inner 
10-15~kpc of the Galaxy and has a typical 
metallicity of [M/H] $\sim$ $-$1.6 (\citealt{car07}; see also \citealt{giz97}).
Membership in the inner halo suggests an origin in the dissipative mergers of satellite galaxies (e.g., \citealt{sea78,chi01,bel08}).
It is possible that the similarity of these orbits arise from selection effects. Stars spend a larger percentage of their orbital periods near apoaps (in this case near the Sun), and short, highly eccentric orbits would more frequently align with the Sun's Galactic position. 
However, the recent discovery of the L subdwarf 2MASS~J0616-6406, whose highly retrograde orbit extends out to $\gtrsim$30~kpc making it a likely member of the Galaxy's outer halo population \citep{cus09}, suggests that ultracool subdwarfs may nevertheless be well-mixed in the vicinity of the Sun.

\section{Atmospheric Properties}

The distance and kinematics of {\namesh}
do not provide useful constraints on its atmospheric properties---{\teff}, surface gravity ({\logg}) and metallicity ([M/H]).
Such determinations require empirical calibrations (e.g., well-characterized coeval companions or cluster properties; \citealt{giz97b}) or
direct comparison to spectral models (e.g., \citealt{woo06}).  As {\namesh}
is a seemingly isolated source, we used the latter approach, employing the most recent generation
of the {\sc Cond-Phoenix} atmosphere
simulations \citep{hau97,hau99,bar03}
and the {\sc Drift-Phoenix} model atmospheres \citep{deh07,hel08b,wit08}.

\subsection{Spectral Models}

{\sc Phoenix} is a general-purpose model atmosphere code, using plane-parallel geometry, thermochemical equilibrium calculations and
opacity sampling to self-consistently solve for the temperature-pressure profile,
chemical abundances and 
radiative/convective energy transfer through the atmosphere.  
Several implementations of {\sc Phoenix} have been used to study late-type dwarfs,
incorporating various assumptions on elemental abundances, atomic and molecular
opacities, line profiles, condensate formation and convective overshoot 
(e.g., \citealt{hau99,all01,joh08,hel08b}).  

Here we examine two implementations of
{\sc Phoenix}, the GAIA-{\sc Cond} models  \citep{hau03} and the {\sc Drift} models
\citep{deh07,hel08b,wit08},
which differ only in their treatment of the dust cloud layers.  The
former are most similar to the {\sc Cond-Phoenix}
model set developed by \citet{all01}, in which condensate species are
treated as element sinks only and phase-equilibrium is assumed. Hence,
{\sc Cond-Phoenix} models simulate dust-free but element-depleted
atmospheres. The {\sc Drift-Phoenix} models 
apply an advanced model of non-equilibrium grain
formation, including seed formation, growth, evaporation,
sedimentation and convective up-mixing
to simulate the size distribution,
abundances and vertical distribution of grains and their material
composition \citep{woi03,woi04,hel06,hel08a}. In this approach, each
size-variable grain is made of a variety of compounds which
changes with height according to its formation history.  Alternate 
prescriptions for modeling condensate grain formation in cool
dwarf atmospheres have been explored
by \citet{ack01,all03,coo03,tsu02,tsu05,tsu04} and \citet{bur06}; a
thorough comparison of these cloud
models is given in \citet{hel08c}.

For this study, we employed both GAIA {\sc Cond-Phoenix} 
and {\sc Drift-Phoenix} atmosphere
models spanning 2000 $\leq$ {\teff} $\leq$ 3500~K (steps of 100~K),
{\logg} = 5.0 and 5.5 (cgs) and $-$3.0 $\leq$ [M/H] $\leq$ 0.0
(steps of 0.5~dex; Solar [M/H] $\equiv$ 0).  Elemental abundances are scaled
from \citet{and89} and \citet{gre92}.  

\subsection{Color Comparisons}

We pursued a two-step comparison of {\namesh} to the models, first
examining optical/near-infrared colors $i^{\prime}-J$ and $J-K_s$ (see
also \citealt{sch1444,dah08,schil09}).  Synthetic colors were computed directly
from the model atmosphere spectra by convolving each with filter profiles from SDSS
\citep[on the AB magnitude system]{fuk96} and 2MASS \citep[on the Vega
magnitude system]{coh03}.  The filter profiles include the effects of
detector quantum efficiency, telescope throughput and atmospheric
transmission, all of which are essential for the complex
spectra of late-type dwarfs (e.g., \citealt{ste04}).

Figure~\ref{fig_color} displays colors for both model sets for {\logg}
= 5.5 and the full range of {\teff} and [M/H].  In addition to
measurements for {\namesh}, we all show color data for five subdwarfs
classified sdM8 and later: the sdM8 LSR 1425+7102, \citep{lep1425};
the sdM8.5 2MASS~J01423153+0523285 \citep[hereafter
2MASS~J0142+0523]{megmos}; SSSPM~J1013-1356, 2MASS~J1626+3925 and
2MASS~J0532+8246.  Near-infrared photometry for all sources are from
2MASS or \citet{schil09}, with the exception of 2MASS~J0142+0523 for which a synthetic $J-K_s$ color was calculated using
near-infrared spectral data from \citet{mewide3}.
SDSS $i^{\prime}$ magnitudes for {\namesh} and 2MASS~J1626+3925 are
from SDSS DR6; for the other sources, $i^{\prime}$ magnitudes were
bootstrapped from $I_N$ photometry from the SuperCosmos Sky Survey (SSS; \citealt{ham01a,ham01b,ham01c}) by calculating synthetic $i^{\prime}-I_N$ colors from published optical
spectral data for these sources
\citep{me0532,lep1425,sch1444,megmos}. The uncertainties for the
spectroscopy-based magnitudes were assumed to be 0.1~mag, based on
prior work (e.g., \citealt{me02a}).

For the GAIA {\sc Cond-Phoenix} models, predicted colors encompass the
measured values of all the subdwarfs shown with the exception of
2MASS~J0532+8246.  Lower temperatures for a given metallicity
generally yield redder $i^{\prime}-J$ and bluer $J-K_s$ colors, except
at for [M/H] $<$ -2.0 for which $i^{\prime}-J$ colors actually turn blue.
Lower metallicities at a given {\teff} lead to bluer
$i^{\prime}-J$ and $J-K_s$ colors.  {\namesh} and 2MASS~J1626+3925 both fall along the [M/H] = -1.0 line in this diagram (as do LSR~1425+7102 and 2MASS~J0142+0523) around {\teff} = 2300 and 2150~K, respectively.   For {\logg} = 5.0, colors for these sources agree with model atmosphere metallicities and temperatures $\sim$0.2~dex and
$\sim$100~K lower, respectively.  The inferred model parameters based on these colors must be treated with
caution, however, as the [M/H] = 0 GAIA {\sc Cond-Phoenix}
models do not track well with mean
$i^{\prime}-J$ versus $J-K_s$ colors of M0--L0 field dwarfs as
compiled by \citet{wes08}.  The divergence is likely due to absence of
condensate opacity in the GAIA models, which are known to be necessary in 
reproducing the colors of late-type M and L dwarfs (e.g., \citealt{all01}).

The {\sc Drift-Phoenix} models exhibit very different color trends
below {\teff} $\approx$ 2500~K as condensates
become a prominent opacity source in the photosphere.  both $i^{\prime}-J$ and $J-K_s$ colors
trend redder for lower {\teff} for [M/H] $>$ -2.5, with a notable kink
in color tracks at {\teff} $\approx$ 2200--2300 not present in
the GAIA {\sc Cond-Phoenix}
models.  The additional reddening places the [M/H] = 0 models
into closer agreement with the mean colors of L0 dwarfs, although they still diverge from M5--M9 dwarfs.  The
$J-K_s$ color reversal effects all metallicities, with the result that none of the models reproduce the measured colors of the L subdwarfs
{\namesh}, 2MASS~J1626+3925 and 2MASS~J0532+8246 ({\logg} = 5.0 models show
similar behavior).  The deviation of the models away from the
measured photometry is likely due to the model $i$-band magnitudes,
which are highly sensitive to the strong molecular opacity present at
these wavelengths (see $\S$~4.3).  A rough extrapolation of the models 
suggests lower metallicities for {\namesh} and 2MASS~J1626+3925, [M/H] $\sim$ -2.0 to -1.5.

\subsection{Spectral Comparisons}

\begin{figure}
\centering
\epsscale{0.65}
\plotone{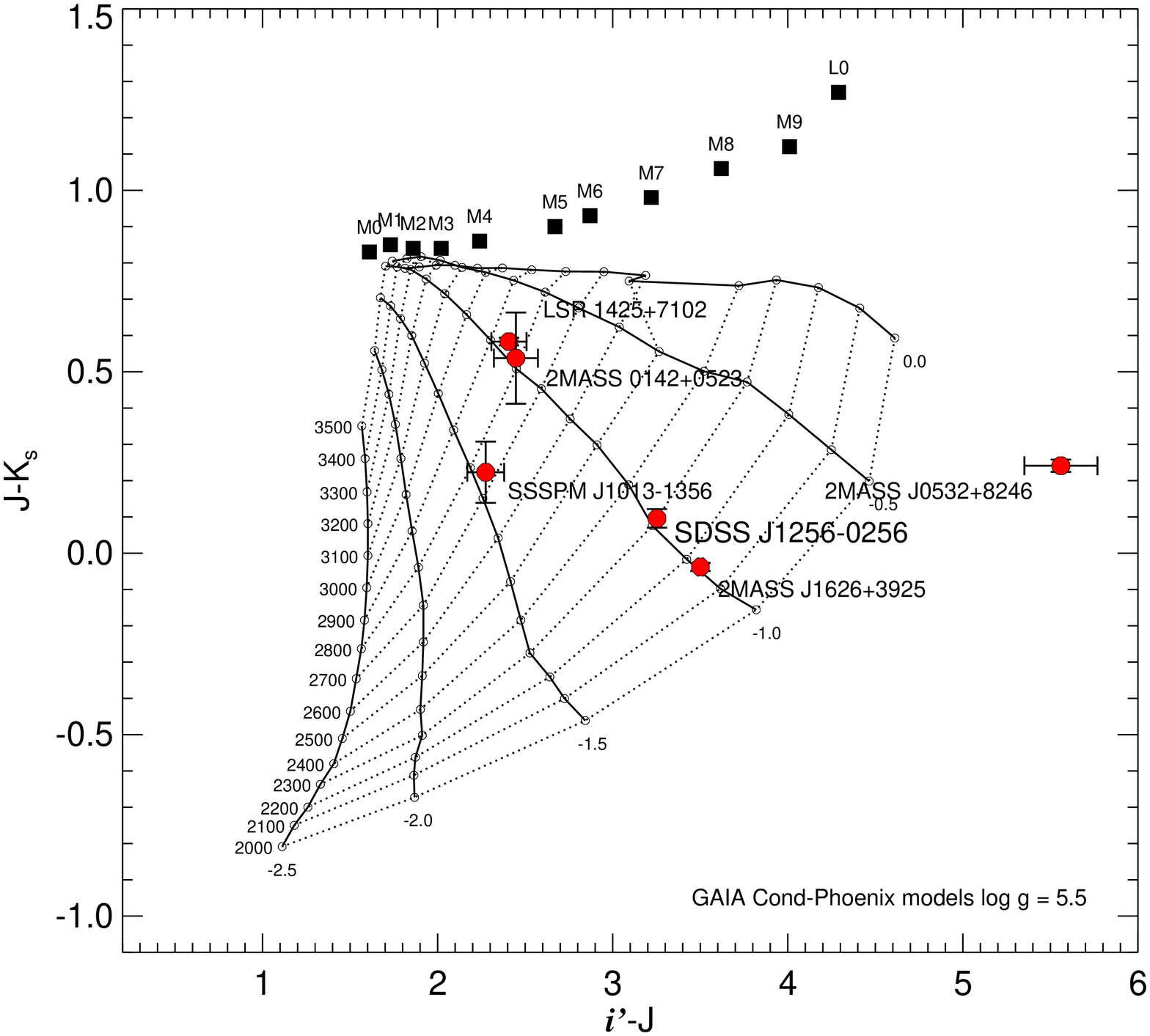}
\plotone{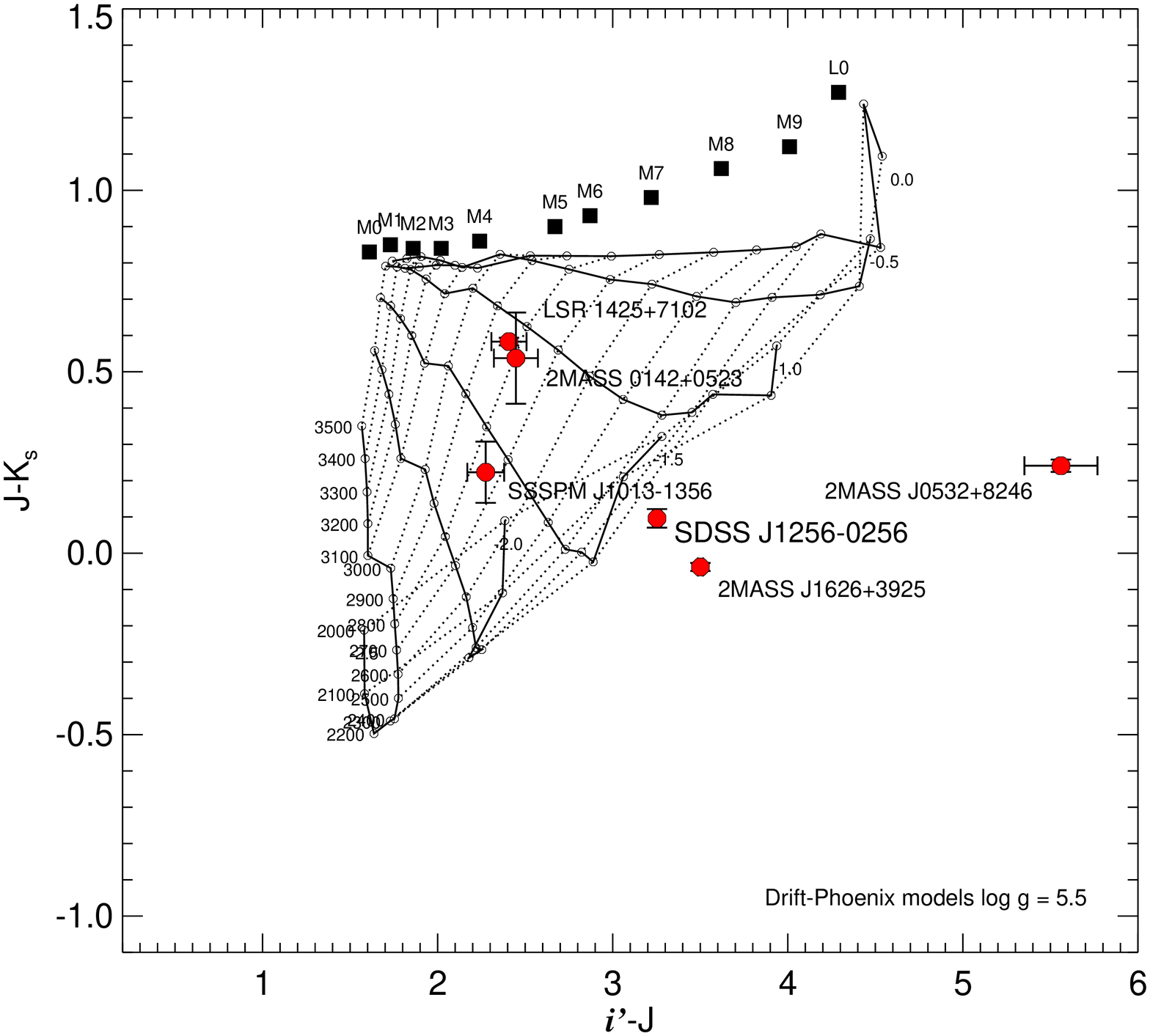}
\caption{SDSS-2MASS $i^{\prime}-J$ versus $J-K_s$ colors for the late-type
subdwarfs 2MASS~J0142+0523, LSR~1424+7102, SSSPM J1013-1356, 
{\namesh}, 2MASS~J1626+3925 and 2MASS~J0532+8246 as compared to 
GAIA {\sc Cond-Phoenix} (top) and {\sc Drift-Phoenix} (bottom) atmospheric model predictions.
Models are shown for {\logg} = 5.5, 
2000 $\leq$ {\teff} $\leq$ 3500~K
in steps of 100~K (along solid lines) and -3.0 $\leq$ [M/H] $\leq$ 0 
in steps of 0.5~dex (along dotted lines).  Also shown are mean SDSS-2MASS
colors of M0--L0 dwarfs from \citet{wes08}.  
\label{fig_color}}
\end{figure}

To further assess the agreement between models with observations, we compared the observed spectrum of {\namesh} directly 
to the GAIA {\sc Cond-Phoenix} and {\sc Drift-Phoenix} model spectra for the same range
of parameters shown in Figure~\ref{fig_color}.  These comparisons 
were made to the combined
red optical and near-infrared spectrum, which was stitched together by first 
smoothing the individual spectra to a common resolution  of {\ldl} = 100,
scaling the spectra to match flux densities in the 0.8--0.9~$\micron$ range, and then 
combining the red optical spectral data for $\lambda < 0.9~\micron$ with
the near-infrared data for $\lambda > 0.9~\micron$\footnote{Red optical data were only used up to 0.9~$\micron$ due to concerns over the relative flux calibration of these data over the 0.9--1.0~$\micron$ range; see discussion in \citet{megmos}.} (see Figure~\ref{fig_nirspec}).
The uncertainty spectrum (flux uncertainty as a function of wavelength) was combined using the same scaling factors and wavelength cutoffs.
We then interpolated the entire spectrum onto a linear
wavelength scale (note that a wavelength scale uniform in frequency
produced similar results).

Observational and model spectra 
were initially normalized over the 0.9--1.0~$\micron$ range;
then, for each model spectrum, a goodness-of-fit statistic was calculated,
\begin{equation}
\Gamma_{\{p\}} = \sum_{\{\lambda\}}{W(\lambda)\frac{[F(\lambda)-{\alpha}S_{\{p\}}(\lambda)]^2}{{\alpha}S_{\{p\}}(\lambda)\sigma(\lambda)}}.
\end{equation}
Here, $F(\lambda)$ and $\sigma(\lambda)$ are the observed
spectrum and uncertainty; $S_{\{p\}}(\lambda)$ is the model spectrum 
for model parameters ${\{p\}}$ = \{{\teff}, {\logg}, [M/H]\}; $\alpha$ is a normalization scale
factor for the model spectrum; and $W(\lambda)$ is a weighting function that
satisfies $\sum_{\{\lambda\}}{W(\lambda)} = 1$.
This form was chosen as a compromise between a standard $\chi^2$ formulation
(e.g., $\sum{[F-S]^2/S}$) and a reduced $\chi^2$ formulation
(e.g., $\sum{[F-S]^2/\sigma^2}$), as the former places too much emphasis 
on the lowest signal regions (i.e., strong absorption features)
while the latter places too much emphasis on the highest signal-to-noise
continuum regions (e.g., the 0.9--1.1~$\micron$ peak).
For alternate approaches, see \citet{tak95} and \citet{cus08}.  
The sums were computed over the spectral
range $\{\lambda\}$ = 0.64--2.4~$\micron$. 
The normalization factor $\alpha$ was allowed to vary over 0.5--1.5
to account for continuum offsets between the observed and model spectra, and the normalization with the minimum $\Gamma$ was retained.  
Various weighting functions $W(\lambda)$ were considered, but ultimately we
settled on one that was constant for all wavelengths.\footnote{\citet{cus08}, in their
analysis of optical and infrared spectra of late-type dwarfs, considered
a weighting function that scaled with the width of each spectral wavelength bin.
As we interpolate the observed and model spectra onto a common, linear wavelength scale, our constant weighting scheme is equivalent.}
Note that we do not consider $\Gamma$ a
robust estimator; it merely provides a quantitative measure of the best-fit model to the data.  The best fits (minimum $\Gamma$) were also visually compared
to verify that they did indeed provide a good match to the
data.

\begin{figure}
\centering
\epsscale{1.05}
\plottwo{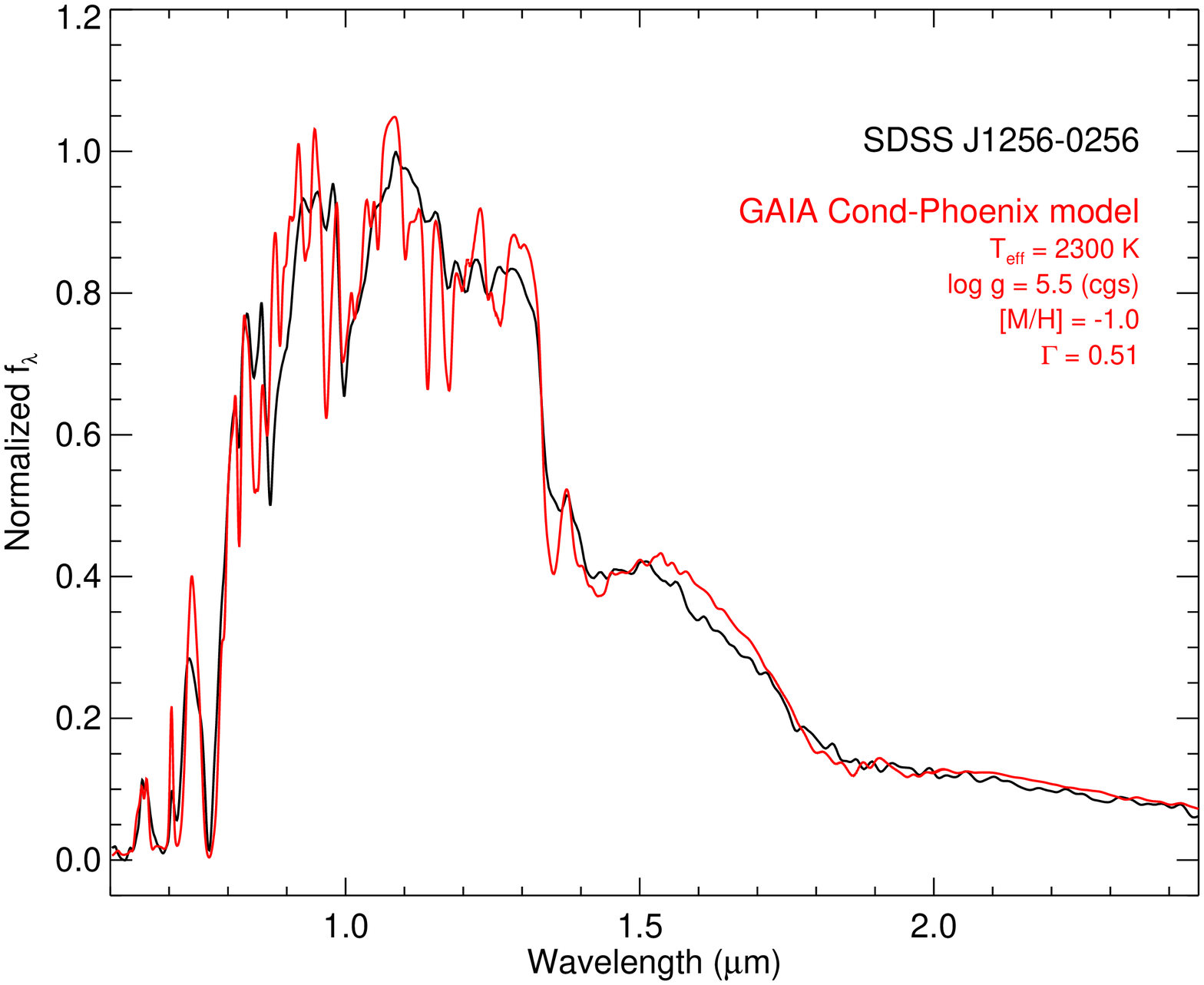}{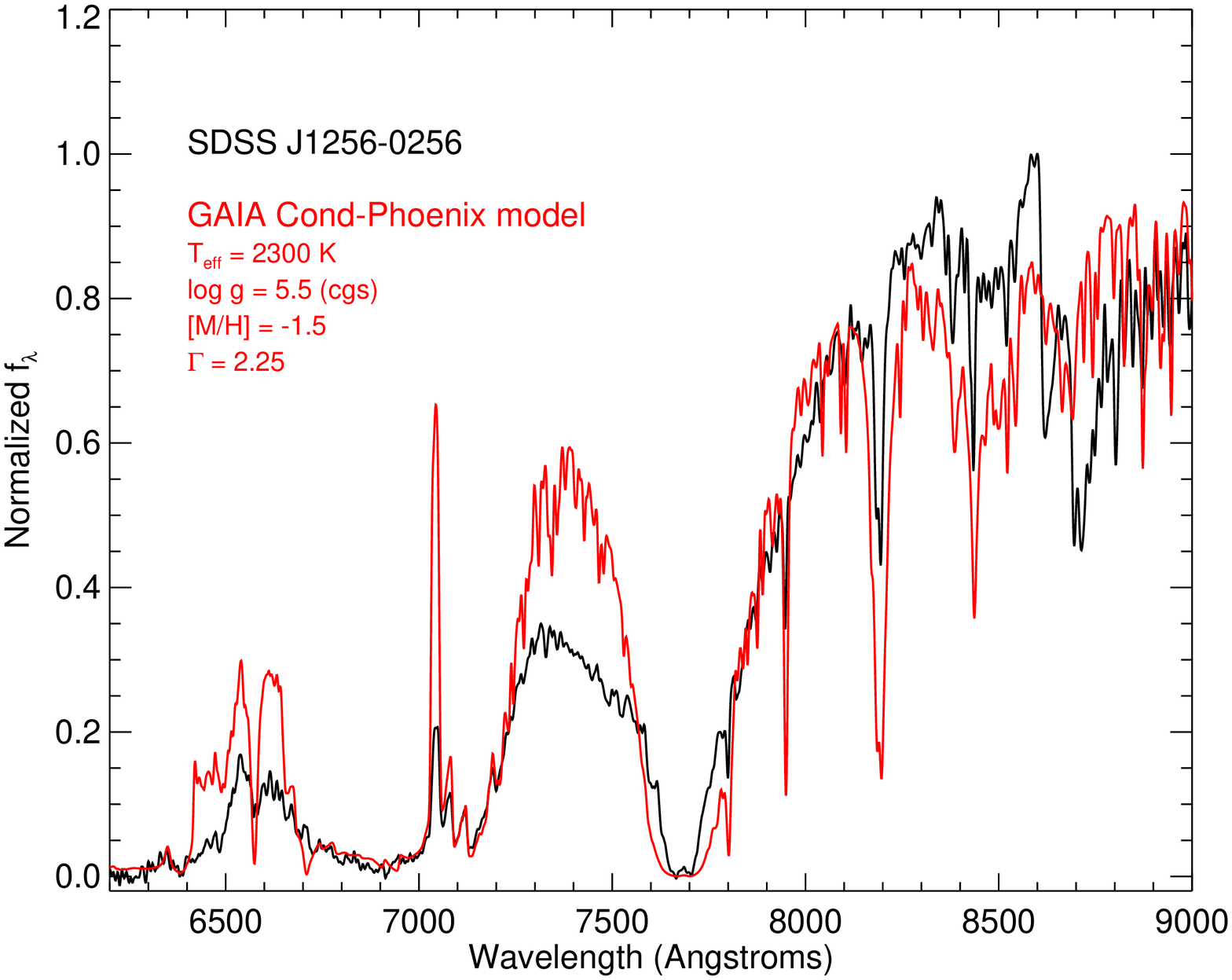}
\plottwo{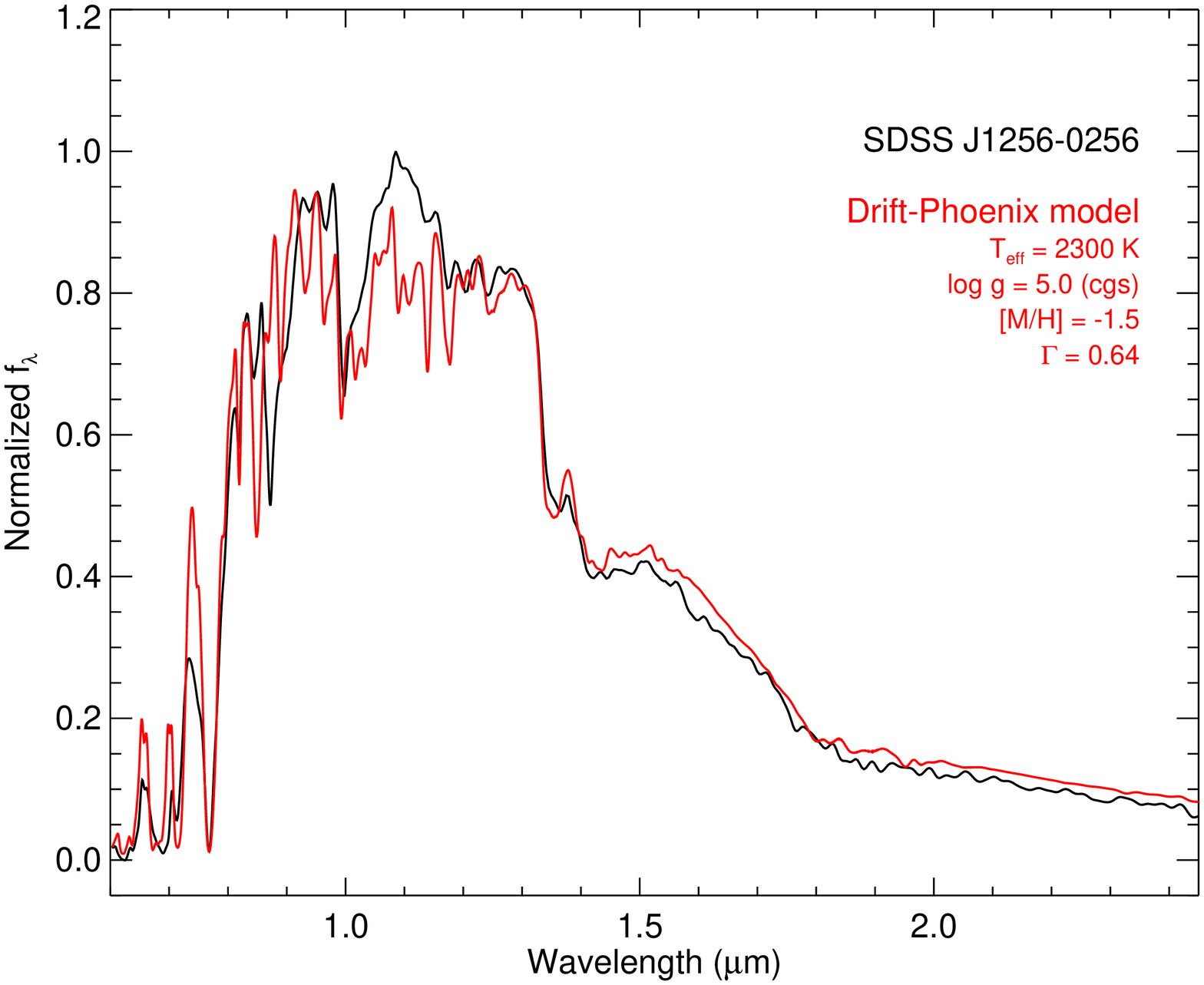}{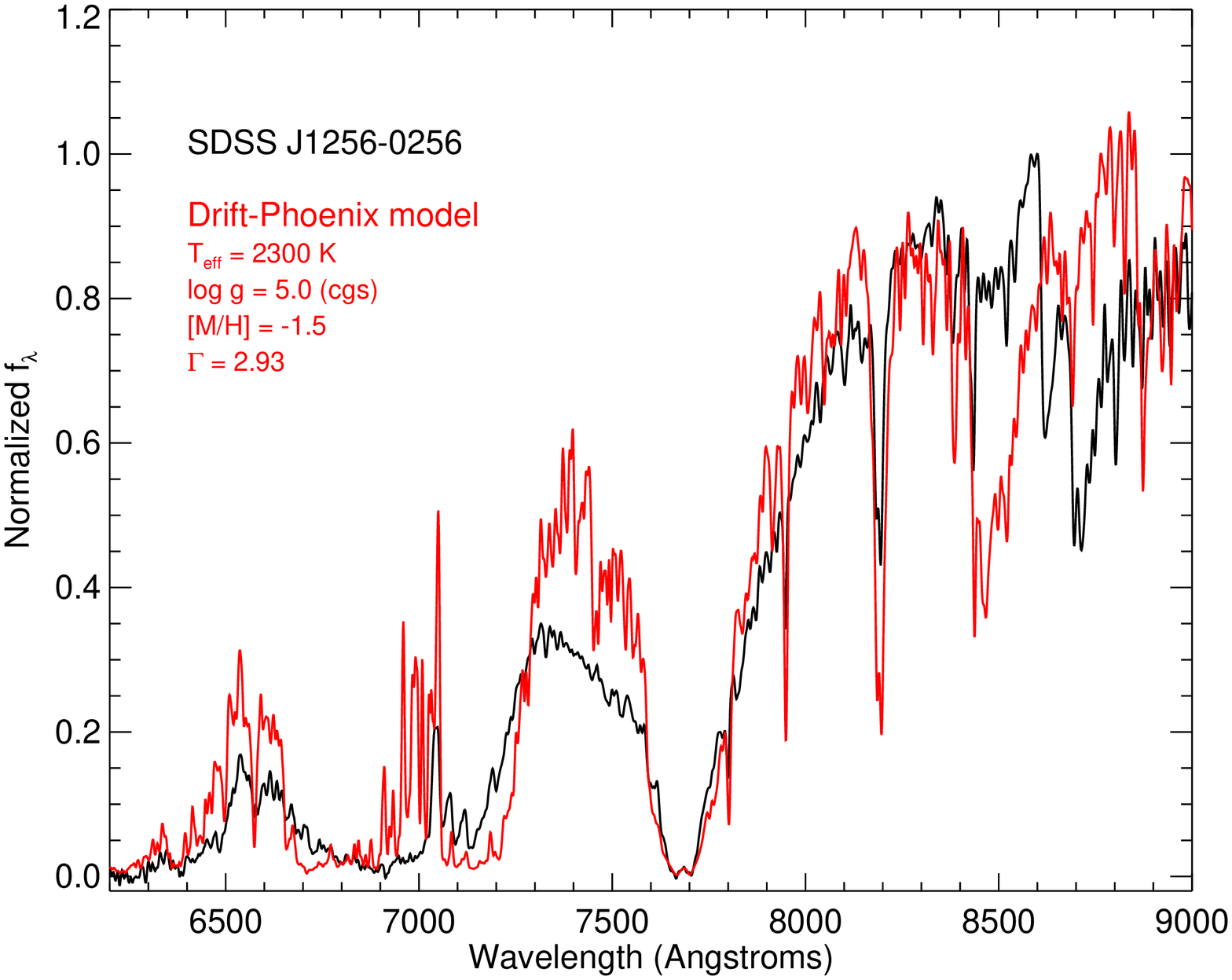}
\caption{(Left) Best-fit GAIA {\sc Cond-Phoenix} (top) and {\sc Drift-Phoenix} models (bottom; red lines) compared with the combined red optical and near-infrared spectrum of  {\namesh} (black lines).  Both models and spectra have been smoothed to a resolution {\ldl} = 100 and interpolated onto a common wavelength scale.  Observational data are normalized in 0.9--1.0~$\micron$ range while models are normalized so as to minimize residuals.  Best-fit parameters are listed in Table~\ref{tab_fit}. (Right) Comparison of best-fit models (based on 
full spectrum) to observed data in the red optical.  Here, both models and spectra have been smoothed to a resolution {\ldl} = 1000 and interpolated onto a common wavelength scale.  The goodness-of-fit
statistic $\Gamma$ computed over the 6400--9000~{\AA} range is notably worse than
that for the full optical/near-infrared spectrum.
\label{fig_fit}}
\end{figure}

Table~\ref{tab_fit} lists the parameters for the five best fits for the 
GAIA {\sc Cond-Phoenix} and {\sc Drift-Phoenix} models; the single best-fit  models 
are compared with the data in Figure~\ref{fig_fit}.  
The best-fit model parameters
are similar between the two sets, with 
{\teff} = 2300--2500~K and [M/H] = -1.5 to -1.0  for the GAIA {\sc Cond-Phoenix} models, and 
{\teff} = 2100--2400~K and  [M/H] = -1.5 and 
 for the {\sc Drift-Phoenix} models.  These parameters includes models with {\logg} = 5.0 and 5.5; lower gravities are generally matched with lower metallicities, which may simply indicate a tradeoff in the inferred photospheric pressure (P$_{ph}$ $\propto$ $g/\kappa$, where $\kappa$ is the Rosseland mean opacity which is generally smaller for lower metallicities). A higher surface gravity is in fact preferred if this object is a low-mass member of the halo population as evolutionary models predict that a 5~Gyr source with {\teff} = 2100--2500~K should have {\logg} = 5.3--5.5 and mass 0.08--0.085~{\msun}  \citep{bur01,bar03}.  Indeed, there are more best-fitting models with {\logg} = 5.5 than 5.0.
 
Encouragingly, the best-fit parameters for both
GAIA {\sc Cond-Phoenix} and {\sc Drift-Phoenix} models
are similar to the best-fit parameters 
from the GAIA {\sc Cond-Phoenix}
color comparison.  The metallicities inferred from both
color and spectral comparisons are also consistent with the mean metallicities of inner halo stars \citep{car07}.  In addition, we find that both sets of models do a
reasonably good job at matching the overall spectral energy
distribution of {\namesh}, in particular fitting the blue spectral
slope from 1.3--2.4~$\micron$ and the depth of the 1.4 and 1.9~$\micron$
{\wat} bands.  The GAIA {\sc Cond-Phoenix} models also provide a fairly
good match to the 0.9--1.3~$\micron$ spectral peak, reproducing
the strong FeH bands at 0.99 and 1.25~$\micron$ but
predicting excessively strong alkali lines in this region.  The {\sc
Drift-Phoenix} models do a
poorer job in this region, failing to reproduce the 1.1~$\micron$
spectral peak and, like the GAIA {\sc Cond-Phoenix} models, exhibiting
excessively strong alkali lines.  

The are more significant deficiencies in the red optical region, however, with both model sets failing to reproduce spectral features in detail, particularly around
the spectral peaks at 6600 and 7400~{\AA}.
The models predict excessively strong {\rbi} and {\nai} alkali
lines, and appear to be missing CrH and FeH opacity in the 8600--8700~{\AA} region. The {\sc Drift-Phoenix} models exhibit excessively strong TiO absorption at 8400~{\AA}, and pronounced discrepancies in the 
6700~{\AA} CaH and 7200~{\AA} TiO bands.  
Surprisingly, the older GAIA {\sc Cond-Phoenix} models
provide overall better fits to the red optical data, although there are still clear problems in alkali line strengths, excessive emission in the pseudo-continua between 6000-7500~{\AA}, and missing CrH and FeH opacity. 
These discrepancies, which influence model $i^{\prime}-J$ color trends (Figure~\ref{fig_color}) 
are probably attributable in part to inadequate treatment of the far-wing line profiles of the pressure-broadened {\nai} and {\ki}
doublets, as more recent opacity calculations have not incorporated in the present models \citep{bur03,joh08}.  
The treatment of dust formation can influence these line profiles due to feedback on the temperature structure \citep{joh08}.
It is therefore promising that the {\ki} line cores and other alkali line strengths are reproduced better in the {\sc Drift-Phoenix} models.
Incompleteness in molecular opacities (e.g., TiO, CrH and FeH) are also likely responsible, a well-known problem 
in modeling the optical spectra of normal L dwarfs (e.g., \citealt{kir99a}).  
We cannot rule out the additional influence of 
elemental composition variations, which are present in subsolar
metallicity stars (e.g., \citealt{edv93,ful00,asp06}) and can influence the overall atmospheric chemical pathways.  
The red optical region clearly remains problematic for low-temperature 
model atmospheres regardless of metallicity (see also \citealt{megmos}), although the incorporation of condensate dust formation appears to be aiding alkali line fits somewhat.

\subsection{Discussion}

While it appears that the inclusion of condensate grain formation as specified by the {\sc Drift-Phoenix} models provides some improvement to the near-infrared colors and alkali line profiles of ultracool dwarfs and subdwarfs, the overall better spectral and color
fits to {\namesh} by the condensate-free GAIA {\sc Cond-Phoenix} models suggests that grain chemistry may be unimportant in metal-poor atmospheres.  
This has been the conclusion of several studies, citing the presence of
enhanced metal-oxide bands, strong lines from refractory species, and blue near-infrared colors as consistent with largely condensate-free photospheres 
\citep{me0532,giz06,rei06,megmos}.  

However, there are problems with this simple interpretation.
It is clear that TiO and VO features weaken from the M subdwarfs to the L subdwarfs, and continue to weaken even through the current L subdwarf sequence (e.g., \citealt{megmos}).  This trend is consistent with the depletion
of refractory gas-phase elements with decreasing temperature. 
In contrast, TiO and VO bands are stronger in later-type M giants where 
low atmospheric pressure inhibits condensation (see \citealt{lod02}). 
Conclusions that gaseous TiO and VO bands are nevertheless 
enhanced may also be biases by simple
equilibrium treatments of condensation chemistry.
\citet{hel08a} have shown that rare-element compounds,
including Ca- and Ti-bearing condensates, never achieve
phase-equilibrium in low-temperature atmospheres; hence,  
the abundances of gas molecules are generally higher
in non-equilibrium cloud models such as {\sc Drift-Phoenix}.
The fact that the metal-oxide bands observed in the spectrum of {\namesh}
are actually {\em weaker} than predicted by the best-fit
{\sc Drift-Phoenix} cloud models suggests that condensate grain formation
may actually be {\em more efficient} in metal-poor atmospheres than these
models predict.  There are some important caveats to this interpretation, however.  Incomplete molecular and pressure-broadened line opacities are clearly an issue for low-temperature atmospheres, affecting in particular pressure-temperature profiles and associated chemistry through the photosphere.  Also, we have only examined one prescription of grain formation in this study; current cloud
models have not yet reached agreement (e.g., \citealt{hel08c}).
It is clear that conclusive statements on the efficiency of condensation in metal-poor atmospheres are premature; models are simply not yet adequate to address this question.

Another consideration, independent of the state of current model atmospheres, is the possibility that L subdwarfs are simply warmer than 
equivalently-classifed L dwarfs.  The {\teff} inferred
for {\namesh}, roughly in the range 2100--2500~K, is comparable to
those for solar-metallicity L0--L2 dwarfs (e.g., \citealt{vrb04}). At these temperatures, 
condensates play a less prominent role in atmospheric
opacity than for cooler/later-type dwarfs.  Simliarly, the sdL7 2MASS~J0532+8246 has been shown
to have a {\teff} = 1730$\pm$90~K, comparable to L4--L5 dwarfs.  This shift is due in large part to the classification methodology proposed by \citet{megmos}.  
Comparison of L dwarf and subdwarf optical spectra in the 7300--9000~{\AA} range
emphasizes the importance of the pressure-broadened 7700~{\AA} {\ki} doublet. This feature is inherently pressure-sensitive, and at a given temperature will be deeper and broader in the higher-pressure, metal-poor (small $\kappa$) photospheres of L subdwarfs.  As such, the apparent persistence of
TiO absorption in L subdwarfs may simply reflect a shift in temperature scale and not condensation abnormalities (see also \citealt{melehpm2-59}).  

Such shifts in inferred quantities like temperature and metallicity have been  seen as an inherent weakness in the classification scheme
of ultracool subdwarfs in general (e.g., \citet{jao08}).  However, the current state of
flux in theoretical models should also emphasize the importance of divorcing classification (a purely empirical exercise) from interpretive physical parameters.  The former is built upon a set of specific standard stars, not models, although standards are as yet in insufficient supply for the L subdwarfs. 
As more examples are uncovered, measurement of their luminosities and {\teff}s , and improved theoretical modeling particularly of optical spectra,
will provide necessary the constraints to calibrate a future classification sequence and enable more robust assessment on condensate grain formation efficiency that is currently possible.

\section{Summary}

We have presented a thorough analysis of the red optical and near-infrared spectrum of the
L subdwarf {\namesh}, originally identified by \citet{siv09} in the SDSS survey.
This source is similar to the sdL4 2MASS~J1626+3925 at both optical and near-infrared wavelengths, 
and we determine an sdL3.5 classification following the preliminary scheme of
\citet{megmos}.  Using the absolute magnitude/spectral type relations for ultracool subdwarfs recently defined by \citet{cus09}, we estimate a distance of
66$\pm$9~pc to this source, formally consistent with the less precise 90$\pm$23~pc parallax distance measurement made by \citet{schil09}.  Combined with its high proper motion and radial velocity, we 
confirm that {\namesh} is a kinematic member of the Galactic inner halo, with a modestly eccentric and highly inclined Galactic orbit whose apoaps is near the Sun.  A comparison of the colors and spectra of {\namesh}
to GAIA {\sc Cond-Phoenix} and {\sc Drift-Phoenix} atmospheric models indicate best-fit atmospheric parameters of 
{\teff} = 2100--2500~K and [M/H] = -1.5 to -1.0 for {\logg} = 5.0--5.5, although discrepancies between the
models and the data, particularly in the red optical region, indicate that these
parameters be treated with caution.  Comparisons to the {\sc Drift-Phoenix} models contradict prior conclusions that condensate
formation may be inhibited in metal-poor, low-temperature atmospheres. Non-equilibrium grain species abundances (particularly for Ti- and Ca-bearing species) predict even stronger metal-oxide bands than those observed in the spectrum of {\namesh}, and thus enhanced condensate formation in the atmosphere of this metal-poor source.  Improvements to the model atmospheres in the red optical region are necessary before any conclusive statement can be made.  In addition, the possibility that the temperature scale of L subdwarfs is warmer than that of L dwarfs may provide a sufficient
explanation for strong metal-oxide bands irrespective of grain chemistry.

\acknowledgments

The authors would like to thank telescope operators Dave Griep
and Mauricio Martinez,
and instrument specialists John Rayner and Jorge Bravo
for their assistance during the IRTF and Magellan observations.
AJB thanks J.\ Bochanki and A.\ West for useful discussion on 
Galactic coordinate systems.
SW acknowledges the Graduierten Kollege 1351 from the German Research Council. 
This publication makes use of data 
from the Two Micron All Sky Survey, which is a
joint project of the University of Massachusetts and the Infrared
Processing and Analysis Center, and funded by the National
Aeronautics and Space Administration and the National Science
Foundation.
2MASS data were obtained from the NASA/IPAC Infrared
Science Archive, which is operated by the Jet Propulsion
Laboratory, California Institute of Technology, under contract
with the National Aeronautics and Space Administration.
Based on observations obtained 6.5 meter 
Magellan Telescopes located at Las Campanas Observatory, Chile.
The authors wish to recognize and acknowledge the 
very significant cultural role and reverence that 
the summit of Mauna Kea has always had within the 
indigenous Hawaiian community.  We are most fortunate 
to have the opportunity to conduct observations from this mountain.

Facilities: \facility{IRTF (SpeX)}, \facility{Magellan Clay (LDSS-3)}

\clearpage


\clearpage

\clearpage

\clearpage

\clearpage

\clearpage

\clearpage

\clearpage

\begin{deluxetable}{lcl}
\tabletypesize{\small}
\tablecaption{Properties of {\name}. \label{tab_properties}}
\tablewidth{0pt}
\tablehead{
\colhead{Parameter} &
\colhead{Value} &
\colhead{Ref.} \\
}
\startdata
$\alpha_{J2000}$ & 12$^h$56$^m$37$\fs$16 & 1 \\
$\delta_{J2000}$ & $-$02$\degr$24$\arcmin$52$\farcs$2 & 1 \\
Spectral Type & sdL3.5 & 2 \\
$i^{\prime}-J$ & 3.253$\pm$0.024 & 3,4 \\
$J-K_s$ & 0.097$\pm$0.026 & 4 \\
$M_{J}$\tablenotemark{a} & 12.23$\pm$0.22 & 2,5 \\
$M_{H}$\tablenotemark{a} & 11.81$\pm$0.23 & 2,5 \\
$M_{K_s}$\tablenotemark{a} & 11.75$\pm$0.25 & 2,5 \\
$d_{est}$ (pc)\tablenotemark{b} & 66$\pm$9 & 2,4,5 \\
$V_{tan}$ ({\kms}) & 186$\pm$26 & 2,4 \\
$V_r$ ({\kms}) & $-$130$\pm$11 & 2 \\
$U$ ({\kms}) & $-$115$\pm$11 & 2 \\
$V$ ({\kms}) & $-$101$\pm$18 & 2 \\
$W$ ({\kms}) & $-$150$\pm$9 & 2 \\
{\teff} (K) & $\sim$ 2100--2500 & 2 \\
{\logg} (cgs) & $\sim$ 5.0--5.5 & 2 \\
{[M/H]} (dex) & $\sim$ -1.5 -- -1.0 & 2 \\
\enddata
\tablenotetext{a}{Estimated absolute magnitudes based on the absolute magnitude/spectral type relations of \citet{cus09} and spectral type sdL3.5.}
\tablenotetext{a}{Note that \citet{schil09} measure an astrometric distance of
90$\pm$23~pc for this source, formally consistent with our more precise
estimate based on the \citet{cus09} absolute magnitude/spectral type relations.}
\tablerefs{(1) 2MASS \citep{skr06}; (2) This paper; (3) SDSS \citep{ade08}; (4) \citet{schil09}; (5) \citet{cus09}.}
\end{deluxetable}

\begin{deluxetable}{lcc}
\tabletypesize{\small}
\tablecaption{Atomic Line Equivalent Widths (EW).\label{tab_ews}}
\tablewidth{0pt}
\tablehead{
\colhead{Feature} &
\colhead{Line Center} &
\colhead{EW} \\
\colhead{} &
\colhead{({\AA})} &
\colhead{({\AA})} \\
}
\startdata
H$\alpha$ & \nodata & $>$-0.9 \\
\ion{Ca}{1} & 6571.10 & 3.1$\pm$0.6  \\
\ion{Ti}{1} & 7204.47 & 2.4$\pm$0.5  \\
\ion{Rb}{1} & 7798.41 & 2.9$\pm$0.4  \\
\ion{Rb}{1} & 7945.83 & 2.6$\pm$0.2  \\
\ion{Na}{1} & 8182.03 & 9.1$\pm$0.2\tablenotemark{a}  \\
\ion{Na}{1} & 8193.33 & \nodata  \\
\ion{Ti}{1} & 8433.31 & 3.7$\pm$0.3  \\
\ion{Cs}{1} & 8519.19 & 1.3$\pm$0.2  \\
\ion{Ca}{2} & 8540.39 & 0.5$\pm$0.3  \\
\ion{Cs}{1} & 8941.47 & 1.1$\pm$0.4  \\
\enddata
\tablenotetext{a}{EW for combined doublet.}
\end{deluxetable}

\begin{deluxetable}{lcccc}
\tabletypesize{\small}
\tablecaption{Spectral Model Fits.\label{tab_fit}}
\tablewidth{0pt}
\tablehead{
\colhead{Model} &
\colhead{\logg} &
\colhead{{\teff}} &
\colhead{[M/H]} &
\colhead{$\Gamma$} \\
 & 
\colhead{(cgs)} &
\colhead{(K)} &
\colhead{(dex)} & \\
}
\startdata
GAIA	 {\sc Cond-Phoenix} & 5.5 & 2300 & -1.0 & 0.51 \\
		& 5.5 & 2400 & -1.0 & 0.53 \\
		& 5.0 & 2400 & -1.5 & 0.56 \\
		& 5.0 & 2300 & -1.5 & 0.57 \\
		& 5.5 & 2500 & -1.0 & 0.72 \\
{\sc Drift-Phoenix}	& 5.0 & 2300 & -1.5 & 0.64 \\
		& 5.0 & 2200 & -1.5 & 0.66 \\
		& 5.5 & 2300 & -1.5 & 0.67 \\
		& 5.5 & 2400 & -1.5 & 0.75 \\
		& 5.5 & 2100 & -1.5 & 0.76 \\
\enddata
\tablecomments{Top five best fit models for combined optical and near-infrared spectrum
of {\namesh}, ranked by $\Gamma$.  Best-fit models are shown in Figure~\ref{fig_fit}. }
\end{deluxetable}

\end{document}